\documentclass[aps,prl,twocolumn]{revtex4-1}

\usepackage{ifthen}

\newcommand{\psec}[1]{\emph{#1.}---}

\newcommand{\bq}{\begin{equation}}
\newcommand{\eq}{\end{equation}}
\newcommand{\bqa}{\begin{eqnarray}}
\newcommand{\eqa}{\end{eqnarray}}

\newcommand{\Sm}{S_{\rm m}}
\newcommand{\rmd}{\ensuremath{\mathrm{d}}}
\newcommand{\delg}{\delta g^{00}}
\newcommand{\Mbar}{\bar{M}}
\newcommand{\Mhat}{\hat{M}}

\newcommand{\dOmega}{\dot{\Omega}}
\newcommand{\ddOmega}{\ddot{\Omega}}
\newcommand{\rhomb}{\bar{\rho}_{\rm m}}
\newcommand{\Pmb}{\bar{P}_{\rm m}}

\newcommand{\Mone}{\mathcal{M}_{\rm I}}
\newcommand{\Mtwo}{\mathcal{M}_{\rm II}}
\newcommand{\Mthree}{\mathcal{M}_{\rm III}}
\newcommand{\Aone}{\mathcal{A}_1}
\newcommand{\Atwo}{\mathcal{A}_2}
\newcommand{\Athree}{\mathcal{A}_3}
\newcommand{\Afour}{\mathcal{A}_4}
\newcommand{\Afive}{\mathcal{A}_5}
\newcommand{\Asix}{\mathcal{A}_6}
\newcommand{\Aseven}{\mathcal{A}_7}
\newcommand{\Aeight}{\mathcal{A}_8}

\begin{document}

\title{Classifying Linearly Shielded Modified Gravity Models in Effective Field Theory}

\author{Lucas~Lombriser}
\affiliation{Institute for Astronomy, University of Edinburgh, Royal Observatory, Blackford Hill, Edinburgh, EH9~3HJ, U.K.}
\author{Andy~Taylor}
\affiliation{Institute for Astronomy, University of Edinburgh, Royal Observatory, Blackford Hill, Edinburgh, EH9~3HJ, U.K.}

\date{\today}

%%%%%%%%%%%%%%%%%%%%%%%%%%%%%%%%%%%%%%%%%%%%%%%%%%%%%%%%%%%%%%%%%%%%%%%%%%%%%

\begin{abstract}

We study the model space generated by the time-dependent operator coefficients in the effective field theory of the cosmological background evolution and perturbations of modified gravity and dark energy models.
We identify three classes of modified gravity models that reduce to Newtonian gravity on the small scales of linear theory.
These general classes contain enough freedom to simultaneously admit a matching of the concordance model background expansion history.
In particular, there exists a large model space that mimics the concordance model on all linear quasistatic subhorizon scales as well as in the background evolution.
Such models also exist when restricting the theory space to operators introduced in Horndeski scalar-tensor gravity.
We emphasize that whereas the partially shielded scenarios might be of interest to study in connection with tensions between large and small scale data, with conventional cosmological probes, the ability to distinguish the fully shielded scenarios from the concordance model on near-horizon scales will remain limited by cosmic variance.
Novel tests of the large-scale structure remedying this deficiency and accounting for the full covariant nature of the alternative gravitational theories, however, might yield further insights on gravity in this regime.

\end{abstract}

%%%%%%%%%%%%%%%%%%%%%%%%%%%%%%%%%%%%%%%%%%%%%%%%%%%%%%%%%%%%%%%%%%%%%%%%%%%%%

\maketitle

%%%%%%%%%%%%%%%%%%%%%%%%%%%%%%%%%%%%%%%%%%%%%%%%%%%%%%%%%%%%%%%%%%%%%%%%%%%%%

\psec{Introduction}
Identifying the nature of the late-time accelerated expansion of our Universe constitutes one of the greatest puzzles of modern science.
Instead of a cosmological constant, cosmic acceleration may be driven by a modification of gravity or the contribution of a dark energy.
A large number of such alternative models have been proposed, and it is, therefore, of great interest to develop a formalism within which they can systematically be explored while providing a consistent description of their cosmological implications.
The past few years have witnessed extensive efforts in the development of such generalized frameworks.
The effective field theory (EFT) of cosmic acceleration~\cite{creminelli:08,gubitosi:12,bloomfield:12} is one such approach and provides a generalized formalism for computing the evolution of the spatially homogeneous and isotropic background of our Universe and the perturbations around it for general modified gravity and dark energy scenarios.
EFT and similar frameworks, however, allow a plethora of imaginative theories of gravity and dark energy, and hence, it becomes of substantial importance to determine what information has to be extracted from the Universe
to be able to exclude alternatives to the concordance model $\Lambda$ cold dark matter ($\Lambda$CDM).
Naturally, this problem is addressed by characterizing observations by the different scales they probe.
The small-scale structure provides the strongest constraints on departures from general relativity (GR) and viable modifications of gravity demand a mechanism that shields the large-scale
modifications in this regime.
There has been intensive research regarding this requirement in recent years and shielding mechanisms, such as the Vainshtein~\cite{vainshtein:72}, chameleon~\cite{khoury:03a} and symmetron~\cite{hinterbichler:10}, or k-mouflage~\cite{babichev:09} effects are well known to hide
these deviations
in the small-scale structure through nonlinear interactions.
In this Letter, we follow a different approach and consider the cancellation of gravitational modifications in the linear sector.
We then propose a classification of the theory space provided by the EFT framework by characterizing theories by their deviations from the concordance model in the linear quasistatic subhorizon perturbations and according to their potential to recover $\Lambda$CDM phenomenology on different scales.
We identify three classes of models that reduce to GR on small linear scales while effectively modifying gravity on large scales.
In general, these
also admit a matching of the
$\Lambda$CDM
expansion history.  
In particular, we find that there exists a large model space that recovers $\Lambda$CDM phenomenology on all linear quasistatic subhorizon scales as well as in the background.
We discuss applications to Horndeski theory and highlight a few of the observational consequences that these models implicate.

%%%%%%%%%%%%%%%%%%%%%%%%%%%%%%%%%%%%%%%%%%%%%%%%%%%%%%%%%%%%%%%%%%%%%%%%%%%%%

\psec{Effective Field Theory}
We consider the action~\cite{bloomfield:12}
\bqa
 S & = & \frac{1}{2\kappa^2} \int \rmd^4x\sqrt{-g} \left[ \phantom{\frac{}{}} \Omega(t) R - 2\Lambda(t) - \Gamma(t) \delg \right. \nonumber\\
 & & + M_2^4(t) (\delg)^2 - \Mbar_1^3(t) \delg \delta K^{\mu}_{\ \mu} - \Mbar_2^2(t)(\delta K^{\mu}_{\ \mu})^2 \nonumber\\
 & & - \Mbar_3^2(t) \delta K^{\mu}_{\ \nu} \delta K^{\nu}_{\ \mu} + \Mhat^2(t) \delg \delta R^{(3)} \nonumber\\
 & & \left. + m_2^2(t)(g^{\mu\nu}+n^{\mu}n^{\nu})\partial_{\mu} g^{00} \partial_{\nu} g^{00} \right] + \Sm\left[\psi_{\rm m};g_{\mu\nu}\right] \label{eq:eftaction}
\eqa
with $\kappa^2\equiv8\pi\,G$ and bare gravitational constant $G$, where we have set the speed of light in vacuum to unity and use a slightly different notation compared to Ref.~\cite{bloomfield:12}.
Eq.~(\ref{eq:eftaction}) describes the cosmological background and perturbations of a generalized theory of gravity with a single scalar field and with the matter sector obeying the weak equivalence principle.
It is written in unitary gauge, i.e., where the time coordinate is chosen in order to absorb the scalar field perturbation in the metric $g_{\mu\nu}$.
Here, $R$ is the Ricci scalar, $K_{\mu\nu}$ is the extrinsic curvature tensor, $R^{(3)}$ is the Ricci scalar of the spatial metric, and $n^{\mu}$ is the normal to surfaces of constant time.
Perturbations, denoted by $\delta$, are obtained from subtraction of the cosmological background.
As we restrict field equations to linear theory, only quadratic terms are considered in Eq.~(\ref{eq:eftaction}).
Furthermore, we only consider the leading order terms in mass and neglect terms that introduce higher-order time derivatives in the Euler-Lagrange equations.

For the following discussion, we define the space of operators with the associated coefficients $\mathcal{P}\equiv\left\{\Omega,\Lambda,\Gamma,M_2,\Mbar_1,\Mbar_2,\Mbar_3,\Mhat,m_2\right\}$, characterizing the effective time-dependent degrees of freedom of the modified gravity and dark energy models described by Eq.~(\ref{eq:eftaction}).
$\Lambda$CDM is represented by $\Omega=1$ and a constant $\Lambda$ with all other operators vanishing.
In quintessence models, we have a minimal coupling $\Omega=1$ with $\Lambda$ and $\Gamma$ playing the role of the scalar field potential and kinetic term.
A nonzero $M_2$ is introduced in k-essence~\cite{armendariz:99}.
Galileon~\cite{nicolis:08} and kinetic braiding~\cite{deffayet:10} models introduce an additional nonzero $\Mbar_1$.
In Horndeski~\cite{horndeski:74,deffayet:11} theory, all operators, except for $m_2$, are used.
A nonzero $m_2$ is introduced in Lorentz covariance violating Ho\v{r}ava-Lifshitz~\cite{horava:09} gravity.
We refer to Refs.~\cite{gubitosi:12,bloomfield:12} for more details on the mapping of modified gravity theories to $\mathcal{P}$.
In addition to the
elements
in $\mathcal{P}$, we also assume a statistically spatially homogeneous and isotropic metric for our Universe, which introduces another free function of time in the form of the scale factor $a(t)$ or the Hubble parameter $H(t)\equiv\dot{a}/a$, where dots denote derivatives with respect to physical time $t$ here and throughout this Letter.
Hence, we define the total function space $\mathcal{Q}\equiv\mathcal{P}\cup\{H(t)\}$ with $\dim(\mathcal{Q})=10$.
Additionally, the metric also introduces a free constant with the spatial curvature $k_0$.
In $\Sm$, we consider a matter-only universe with pressureless dust $\Pmb=0$ and matter density
$\rhomb = \bar{\rho}_{{\rm m}0} a^{-3}$, which adds another free constant to the model space.

%****Background****%
%
Following Refs.~\cite{gubitosi:12,bloomfield:12}, the background evolution equations associated with the action Eq.~(\ref{eq:eftaction}) become
\bqa
  H^2 + \frac{k_0}{a^2} + H \frac{\dOmega}{\Omega} & = & \frac{\kappa^2\rhomb + \Lambda + \Gamma}{3\Omega}, \label{eq:friedmann1} \\
 3H^2 + 2\dot{H} + \frac{k_0}{a^2} + \frac{\ddOmega}{\Omega} + 2H\frac{\dOmega}{\Omega} & = & \frac{\Lambda - \kappa^2\Pmb}{\Omega}. \label{eq:friedmann2}
\eqa
These equations fix two elements in $\mathcal{Q}$ such that our base EFT model consists of eight free functions of time and two free constants.
In the remainder of this Letter, we will assume the spatial curvature and
present matter density to be specified.

%****Quasistatic sub-horizon limit****%
%
Next, we consider the linear cosmological perturbations in the quasistatic subhorizon limit, i.e., where we neglect time derivatives of the fields and have $k/a \gg H$, respectively.
We will also refer to this regime as the Newtonian limit of the models.
The perturbed modified Einstein equations in EFT have been derived in Ref.~\cite{gubitosi:12,bloomfield:12} and shall not be given here again.
The modifications introduced in these perturbations can be described by two time and scale dependent effective quantities that modify the Poisson equation and introduce a gravitational slip
\bqa
 \mu(a,k) \equiv -\frac{2k^2\Psi}{\kappa^2 \rhomb a^2 \Delta} & = & \frac{g_1\frac{k^2}{a^2} + g_2 + g_3\frac{a^2}{k^2}}{g_4\frac{k^2}{a^2} + g_5 + g_6\frac{a^2}{k^2}}, \\
 \gamma(a,k) \equiv -\frac{\Phi}{\Psi} & = & \frac{g_7\frac{k^2}{a^2} + g_8 + g_9\frac{a^2}{k^2}}{g_1\frac{k^2}{a^2} + g_2 + g_3\frac{a^2}{k^2}},
\eqa
respectively, where $\Psi=\delta g_{00}/2g_{00}$ and $\Phi=\delta g_{ii}/2g_{ii}$.
The time-dependent functions $\{g_{\rm i}(a)|{\rm i}=1,2,\ldots,9\}$ are determined by the
elements
in $\mathcal{Q}$ and can be related to them using, in Ref.~\cite{bloomfield:12}, Eqs.~(4.15) and (4.18) and Table~3, where we describe the gravitational slip through $\gamma=(\varpi+1)^{-1}$ (cf.~Ref.~\cite{gleyzes:13}).
GR, really referring to its Newtonian limit, is recovered when $\mu(a,k)=\gamma(a,k)=1$.
It is important to note that all
functions
in $\mathcal{Q}$, except for $M_2$, appear in the background evolution and quasistatic subhorizon linear perturbation equations.
Hence, $M_2$ will always remain unconstrained in the models discussed in the following and only appears when dealing with the full spatially covariant nature of Eq.~(\ref{eq:eftaction}).

%%%%%%%%%%%%%%%%%%%%%%%%%%%%%%%%%%%%%%%%%%%%%%%%%%%%%%%%%%%%%%%%%%%%%%%%%%%%%

\psec{Linearly shielded models}
We now identify the three classes of models in which, in the formal limit of $k\rightarrow\infty$, $\mu(a,k)$ and $\gamma(a,k)$ become unity, and thus recover GR.
\begin{itemize}
 \item[$\Mone$:] If $g_1$, $g_4$, and $g_7$ are nonzero, GR can be %recovered
 restored in the high-$k$ limit by requiring that $g_1/g_4 = g_7/g_1 = 1$.
 Note that with the dependence of $g_{\rm i}$ on the EFT functions, this also implies that $g_3/g_6 = g_9/g_3 = 1$.
 \item[$\Mtwo$:] In the case that $g_1=g_4=g_7=0$, we must require $g_2/g_5 = g_8/g_2 = 1$ to %restore
 recover GR on small scales, defining a second class of linearly shielded modified gravity models.
 \item[$\Mthree$:] Finally, we have all
models in which
$\mu(a,k)=\gamma(a,k)=1$ on all quasistatic subhorizon scales.
\end{itemize}
To classify the model space of $\mathcal{M}_{\rm I - III}$, we formulate eight conditions $\alpha_{1-8}$ on the operator coefficients in $\mathcal{Q}$:
\bqa
 \alpha_1 & : & m_2^2 = \frac{1}{2}(\Omega-1), \\
 \alpha_2 & : & \Mhat = 0, \\
 \alpha_3 & : & \Mbar_3^2 = -\Mbar_2^2, \\
 \alpha_4 & : & \Mbar_1^3 = \dot{\Omega} - H (\Mbar_3^2+3\Mbar_2^2), \\
 \alpha_5 & : & \Mbar_1^3 = -\dot{\Omega} + 2H \Mbar_3^2, \\
 \alpha_6 & : & \partial_t \Mbar_3^2 = \dot{\Omega} - H \Mbar_3^2 \nonumber\\
 & & +\Mhat^2\frac{\dot{\Omega}+\Mbar_1^3-2H(\Mbar_3^2+2\Mhat^2)}{\Omega-1-2m_2^2+2\Mhat^2}, \\
 \alpha_7 & : & \Gamma = \partial_t\Mbar_1^3 - 4H\partial_t\Mhat^2 + H\Mbar_1^3-4\dot{H}\Mbar_3^2 \nonumber\\
 & & + \frac{2k_0}{a^2}\Mbar_3^2 - 4\dot{H}\Mhat^2 - 4H^2\Mhat^2 \nonumber\\
 & & - ( 4\partial_t\Mhat^2 + \dot{\Omega}-\Mbar_1^3+2H\Mbar_3^2+8H\Mhat^2 ) \nonumber\\
 & & \times\frac{ \dot{\Omega} + \Mbar_1^3 -2H(\Mbar_3^2+2\Mhat^2) }{2(\Omega-1-2m_2^2+2\Mhat^2)}, \\
 \alpha_8 & : & \Mbar_1^3 = \frac{1}{6(\ddot{H}+3H\dot{H})} \left\{\vphantom{\frac{k_0}{a^2}}\left[ -\dot{\Omega}\dot{R}^{(0)} \right. \right. \nonumber\\
 & & \left. - 6\dot{H}(-\Gamma+\partial_t\Mbar_1^3 + 3\dot{H}\Mbar_2^2+\dot{H}\Mbar_3^2)\vphantom{\dot{R}^{(0)}}\right] \nonumber\\
 & & \left. + 24\frac{k_0}{a^2}[H\partial_t\Mhat^2 + (\dot{H}+H^2)\Mhat^2]\right\}.
\eqa
Here, $R^{(0)}$ denotes the background Ricci scalar.
The combination of $\alpha_1$ and $\alpha_2$ sets $g_1/g_4=g_7/g_1=g_3/g_6=g_9/g_3=1$ and $\alpha_3$ sets $g_1=g_4=g_7=0$.
Furthermore, $\alpha_8$ sets $g_3=g_6=g_9=0$.
Supplementing these conditions with combinations of $\alpha_{5-7}$ sets $g_2/g_5=g_8/g_2=1$.
Note that we do not need to formulate a condition for $g_2=g_5=g_8=0$, in which case we must impose $\alpha_1$ and $\alpha_2$.
The corresponding models are part of $\Mthree$ and are contained in a smaller number of combinations of $\alpha_{1-8}$.
To simplify the notation, we define the model spaces that satisfy these conditions as $\mathcal{A}_{\rm i}\equiv\left\{\left.\mathcal{Q}\right|\alpha_{\rm i}\right\}$.
The model classes $\mathcal{M}_{\rm I - III}$ can then be defined as
\bqa
 \Mone & \equiv & (\Aone\cap\Atwo)\backslash\Athree, \\
 \Mtwo & \equiv & \Athree\cap\Asix\cap\Aseven, \\
 \Mthree & \equiv & (\Aone\cap\Atwo\cap\Afour)\cup(\Aone\cap\Atwo\cap\Afive\cap\Asix) \nonumber\\
 &  & \cup(\Athree\cap\Asix\cap\Aseven\cap\Aeight). \label{eq:Mthree}
\eqa
In $\Mone$ and $\Mtwo$, $\mathcal{Q}$ reduces to a six- and five-dimensional space of free functions, respectively, while for $\Mone$, we require $\Mbar_3^2\neq-\Mbar_2^2$.
An interesting subclass of $\Mtwo$ is $\Mtwo\cap\Atwo$ with $g_9/g_3=1$ whereas $g_3/g_6\neq1$ in general.
In $\Mthree$, we have $\Aone\cap\Atwo\cap\Afour$ with $\dim(\mathcal{Q})=5$, as well as $\Aone\cap\Atwo\cap\Afive\cap\Asix$ and $\Mtwo\cap\Aeight$ with $\dim(\mathcal{Q})=4$.

%%%%%%%%%%%%%%%%%%%%%%%%%%%%%%%%%%%%%%%%%%%%%%%%%%%%%%%%%%%%%%%%%%%%%%%%%%%%%

\psec{Concordance model background}
In addition to recovering GR on small scales, we may require that our models mimic a $\Lambda$CDM background expansion history.
The two Friedmann equations, Eqs.~(\ref{eq:friedmann1}) and (\ref{eq:friedmann2}), constrain two
background functions $\{\Omega,\Lambda,\Gamma,H\}$.
As long as the desired combinations of $\alpha_{1-8}$ leave three of these functions free, we can set
$H^2=H_{\Lambda{\rm CDM}}^2 \equiv (\kappa^2\bar{\rho}_{{\rm m}0}a^{-3} + \Lambda_0)/3 - k_0/a^2$, where $\Lambda_0$ denotes the cosmological constant.
Hence, for general $\mathcal{M}_{\rm I-III}$ models, we have sufficient freedom in $\mathcal{Q}$ to make this choice.
In $\Mthree$, now both the Newtonian linear perturbations and the background expansion history match the $\Lambda$CDM phenomenology, while we can have up to four free functions in $\mathcal{Q}$.

%%%%%%%%%%%%%%%%%%%%%%%%%%%%%%%%%%%%%%%%%%%%%%%%%%%%%%%%%%%%%%%%%%%%%%%%%%%%%

\psec{Horndeski scalar-tensor theory}
Having established and classified the linearly shielded modified gravity models in EFT, we now shortly discuss an application of our results to Horndeski theory.
The Horndeski action~\cite{horndeski:74,deffayet:11} describes the most general local, Lorentz-covariant, and four-dimensional scalar-tensor theory for which the Euler-Lagrange equations involve at most second-order derivatives of the scalar and tensor fields.
This restricts the model space with the three conditions $2\Mhat^2 = -\Mbar_3^2 = \Mbar_2^2$ and $m_2 = 0$~\cite{gleyzes:13,bloomfield:13}, reducing $\mathcal{Q}$ to five free functions of time, where we additionally have the two free constants from spatial curvature and present matter density (cf.~Ref.~\cite{bellini:14}).
Hereby, $\alpha_3$ is always satisfied, yielding $g_1=g_4=g_7=0$ (cf.~Ref.~\cite{defelice:11}) and implying that there are no $\Mone$ models in Horndeski theory.
Models of the type $\Mtwo$, on the contrary, can be formulated, for which now $\dim(\mathcal{Q})=3$, where $M_2$ constitutes one of the free functions.
The general $\Mtwo$ class in Horndeski theory also admits a $\Lambda$CDM background.
For $(\Mtwo\cap\Aeight)\subset\Mthree$
with $\alpha_3$ already satisfied by the Horndeski conditions, we obtain $\dim(\mathcal{Q})=2$.
$M_2$ being free, we can choose $H$ as the second free function, which may be set to $H_{\Lambda{\rm CDM}}$.
The other functions then form a closed system of coupled nonlinear differential equations.
The remaining two $\Mthree$ models,
joined with the Horndeski conditions, imply that $\Omega=1$ and $\Mbar_1 = \Mbar_2 = \Mbar_3 = \Mhat = m_2 = 0$, where one condition becomes obsolete in the latter combination.
As $\Omega$ is fixed, the Friedmann equations constrain two
functions in $\{\Lambda,\Gamma,H\}$ with the second free function
being $M_2$.
If furthermore setting $M_2=0$, this reduces to the phenomenology of a quintessence model.
The choice of
$H=H_{\Lambda{\rm CDM}}$
sets $\Gamma = 0$, which when $M_2=0$ corresponds to the EFT of the concordance model.
The scenario $\{\Gamma=0,M_2\neq0\}$ may, in principle, be realized in k-essence.
Note, however, that while the Lagrangian densities in the Horndeski action can be identified which yield the desired phenomenology, the reconstruction is not unique as it is limited to quantities in the background and linear perturbations and may yield somewhat contrived models.

%%%%%%%%%%%%%%%%%%%%%%%%%%%%%%%%%%%%%%%%%%%%%%%%%%%%%%%%%%%%%%%%%%%%%%%%%%%%%

\psec{Observational implications}
$\mathcal{M}_{\rm I-III}$ models provide distinctive features to study in connection with tensions between large and small scale observations, for instance, reported by Planck in Ref.~\cite{planck15:13}.
Moreover, an enhancement of gravity on large scales with consequent reduction of the integrated Sachs-Wolfe effect may also allow a larger contribution of gravitational waves in the low multipoles of the temperature anisotropy power spectrum.
While the models introduce novel and interesting deviations in $\mu(a,k)$ and $\gamma(a,k)$, $\Mthree$ models, behaving as GR on all linear quasistatic subhorizon scales and, in addition, having the ability to mimic $H_{\Lambda{\rm CDM}}$, pose new challenges to the aim of discriminating between the concordance model and more complex scenarios.
The problem becomes apparent with the inclusion of $M_2$, which does not have any impact on the background equations nor on the Newtonian linear perturbations.
In these scenarios, limited constraints may be inferred from observations that test beyond the Newtonian regime such as measurements of the low multipoles of the cosmic microwave background, which can be computed for $\mathcal{M}_{\rm I-III}$ using an existing EFT implementation in a Boltzmann linear theory solver~\cite{hu:13b}.
Note, however, that the near-horizon observations are weakened by cosmic variance, which limits possible constraints on these deviations.
A novel observational method that may remedy this deficiency is the multitracer proposal of Ref.~\cite{seljak:08}.
By dividing galaxies in a galaxy-redshift survey into differently biased samples, combining observations at different modes, and cross correlating the data with weak gravitational shear fields, one obtains a measurement of the growth
of matter density
fluctuations that, in principle, is free of cosmic variance.
This idea has been applied in Ref.~\cite{lombriser:13a} (also see Ref.~\cite{motta:13}) to forecast constraints on
modified gravity and dark energy models using deviations introduced in the relativistic contributions to galaxy clustering.
Thereby, large-scale modifications with high-$k$ suppression have been proposed and shown to provide measurable signatures in future galaxy-redshift surveys.
The $\mathcal{M}_{\rm I-III}$ classes introduced here provide a thorough theoretical motivation for the \emph{ad hoc} phenomenological modifications studied in Ref.~\cite{lombriser:13a}.

%%%%%%%%%%%%%%%%%%%%%%%%%%%%%%%%%%%%%%%%%%%%%%%%%%%%%%%%%%%%%%%%%%%%%%%%%%%%%

\psec{Caveats}
Importantly, when constructing linearly shielded models with the operator coefficients that remain free after applying combinations of $\alpha_{1-8}$ and optionally, a matching of $H_{\Lambda{\rm CDM}}$,
the free coefficients should be designed such that ghost and gradient instabilities can be avoided, ensuring a positive kinetic term and preventing imaginary sound speeds.
Likewise, the feasibility of linear cancellation should be ensured with the applicability of the quasistatic subhorizon approximation, implying a sufficiently large sound speed of the dark energy contribution.
Corresponding inequality conditions on the EFT functions can be inferred from Refs.~\cite{creminelli:08,bloomfield:12}.
Note that when relating the $g_{\rm i}$ terms in $\mu$ and $\gamma$ to the operator coefficients, time derivatives in the perturbed field equations were neglected prior to their combination.
Ref.~\cite{bellini:14} points out that a combination preceding this approximation may change expressions for the high-$k$ subdominant $g_{\rm i}$.
Linear cancellation in $\mathcal{M}_{\rm I-II}$ is unaffected, depending only on the dominant $g_{\rm i}$, whereas $\alpha_4$, $\alpha_5$, $\alpha_8$, defining $\Mthree$, may change.
However, the conclusion that computations need to be performed beyond the quasistatic approximation to test $\Mthree$ is unaltered.
Furthermore, note that while the $\mathcal{M}_{\rm I-III}$ models provide a mechanism for recovering GR on small linear scales, in the limit of $k\rightarrow\infty$, linear theory eventually fails and nonlinearities need to be taken into account, which may introduce new deviations from GR, particularly for relativistic objects.
The scalar degree of freedom may, for instance, introduce modifications in the Solar System, in the decay of the orbital periods in binary-pulsar systems~\cite{yagi:13a}, or in the propagation of gravitational waves from cosmological sources~\cite{saltas:14} (see Ref.~\cite{bellini:14}).
In general, however, the inclusion of nonlinear scales introduces further EFT operators and increases the number of free coefficients in the formalism.
We expect that similar conditions to $\alpha_{1-8}$ can be formulated for these extra coefficients that ensure a recovery of GR on increasingly nonlinear scales.
Ultimately, those may allow the full reconstruction of a corresponding scalar-tensor action in an expansion series around the cosmological background.
Alternatively, linear cancellation may be invoked in nonlinearly shielded models to comply with observations of the cosmic structure.
Models that restore GR on linear scales but allow deviations in the weakly nonlinear regime may also be of interest~\cite{lombriser:14}.
In this context, we emphasize that in Eq.~(\ref{eq:eftaction}) we have focused on terms with leading scaling dimensions, dominating at low energies.
By allowing higher-order spatial derivatives on the perturbed quantities in the action with increasing scaling dimensions, we can obtain additional terms in
$\mu$ and $\gamma$ with higher powers in $k$, imposing an extension to our classification.
However, those become important at increasingly smaller scales, where we also expect our linear computations to fail, and, hence, have been neglected here.
We leave an exploration of
implications of stability requirements on
model building as well as
effects of higher-order derivatives and nonlinearities for future work.

%%%%%%%%%%%%%%%%%%%%%%%%%%%%%%%%%%%%%%%%%%%%%%%%%%%%%%%%%%%%%%%%%%%%%%%%%%%%%

\psec{Discussion}
The past few years have seen extensive efforts in the development of a generalized framework for
describing the cosmological structure formed in modified gravity and dark energy models.
One such formalism is the effective field theory of cosmic acceleration for gravity theories with a single scalar field used here.
We propose a classification of the theory space in the EFT framework by characterizing alternative models by their deviations from the concordance model in the quasistatic subhorizon linear perturbations and background evolution equations, and according to their potential to recover $\Lambda$CDM phenomenology.
We identify three model classes that introduce a cancellation of their gravitational modifications on small linear scales.
The EFT formalism also provides sufficient freedom to impose a $\Lambda$CDM expansion history.
In particular, there exists a large model space that matches $\Lambda$CDM phenomenology in all Newtonian linear perturbations as well as in the background.
Such models also exist when restricting the theory space to operators introduced in Horndeski scalar-tensor gravity.
We emphasize that whereas partially shielded models may be of interest to address tensions between large and small scale observations, the potential of conventional cosmological probes to distinguish the fully shielded scenarios from $\Lambda$CDM is limited by cosmic variance.
Hence, new observational methods need to be developed to remedy this deficiency.
Clearly, those
will have to account for the fully covariant nature of the alternative theories.
One such method that may cast new insights on gravity and dark energy in the near-horizon regime could be the measurement of the relativistic contributions to galaxy clustering using a multitracer analysis of galaxy-redshift surveys to cancel cosmic variance.
This may improve constraints on the theory space over what can be inferred from cosmic microwave background data.
We leave an exploration of such constraints for future work.

%
%

%%%%%%%%%%%%%%%%%%%%%%%%%%%%%%%%%%%%%%%%%%%%%%%%%%%%%%%%%%%%%%%%%%%%%%%%%%%%%

\begin{acknowledgments}
L.L.~thanks Kazuya Koyama for useful discussions.
This work has been supported by the U.K. STFC Consolidated Grant for Astronomy and Astrophysics at the University of Edinburgh.
Please contact the authors for access to research materials.
\end{acknowledgments}

%%%%%%%%%%%%%%%%%%%%%%%%%%%%%%%%%%%%%%%%%%%%%%%%%%%%%%%%%%%%%%%%%%%%%%%%%%%%%

\vfill
\bibliographystyle{arxiv_physrev_mod}
\bibliography{linearshielding}

\end{document}